\documentclass[
reprint,           
superscriptaddress,
amsmath,           
amssymb,           
prd,               
showpacs,          
notitlepage,       
longbibliography,  
floatfix,          
]{revtex4-1}

\usepackage{tensor}     
\usepackage{graphicx}   
\usepackage[
colorlinks=true,        
citecolor=blue,         
linkcolor=blue,         
urlcolor=blue           
]{hyperref}             
\usepackage{bm}         
\usepackage{xcolor}     

\newcommand{\newc}{\newcommand*}
\newc{\figurewidth}{3.2in}
\newc{\xbar}{\bar{x}}
\newc{\rhoeq}{\rho_{\rm{eq}}}
\newc{\zeq}{z_{\rm{eq}}}
\newc{\la}{\lambda}
\newc{\tla}{\tilde{\la}}
\newc{\dt}{\delta}
\newc{\Dt}{\Delta}
\newc{\vj}{\vec{j}}
\newc{\vl}{\vec{l}}
\newc{\hx}{\hat{x}}
\newc{\hy}{\hat{y}}
\newc{\bj}{\bm{j}}
\newc{\mJ}{\mathcal{J}}
\newc{\mP}{\mathcal{P}}
\newc{\ga}{\gamma}
\newc{\Msun}{M_\odot}
\newc{\app}{\approx}
\newc{\av}[1]{\langle #1 \rangle}
\newc{\eq}[1]{Eq.~\eqref{#1}}
\newc{\al}{\alpha}
\newc{\Xstar}{X_{\ast}}
\newc{\seq}{\sigma_{\rm{eq}}}
\newc{\fpbh}{f_{\rm{pbh}}}
\newc{\RR}{{\cal R}}

\def\p{\partial}

\def\({\left(}
\def\){\right)}
\def\[{\left[}
\def\]{\right]}

\def\e{\begin{equation}}
\def\q{\end{equation}}
\def\m{\begin{eqnarray}}
\def\n{\end{eqnarray}}

\begin{document}

\title{Effects of the merger history on the merger rate density of primordial black hole binaries}

\author{Lang Liu}
\email{liulang@itp.ac.cn} 
\affiliation{CAS Key Laboratory of Theoretical Physics,
Institute of Theoretical Physics, Chinese Academy of Sciences,
Beijing 100190, China}
\affiliation{School of Physical Sciences,
University of Chinese Academy of Sciences,
No. 19A Yuquan Road, Beijing 100049, China}
\author{Zong-Kuan Guo}
\email{guozk@itp.ac.cn}
\affiliation{CAS Key Laboratory of Theoretical Physics,
Institute of Theoretical Physics, Chinese Academy of Sciences,
Beijing 100190, China}
\affiliation{School of Physical Sciences,
University of Chinese Academy of Sciences,
No. 19A Yuquan Road, Beijing 100049, China}
\author{Rong-Gen Cai}
\email{cairg@itp.ac.cn}
\affiliation{CAS Key Laboratory of Theoretical Physics,
Institute of Theoretical Physics, Chinese Academy of Sciences,
Beijing 100190, China}
\affiliation{School of Physical Sciences,
University of Chinese Academy of Sciences,
No. 19A Yuquan Road, Beijing 100049, China}
\date{\today}
\begin{abstract}
We develop a formalism to calculate the merger rate density of primordial black hole binaries with a general mass function,
by taking into account the merger history of primordial black holes.
We apply the formalism to three specific mass functions, monochromatic, power-law and log-normal cases.
In the former case, the merger rate density is dominated by the single-merger events,
while in the latter two cases,
the contribution of the multiple-merger events on the merger rate density can not be ignored.
The effects of the merger history on the merger rate density depend on the mass function.
\end{abstract}

\maketitle

\section{Introduction}
Various astrophysical and cosmological observations provide substantial evidences
firmly establishing the existence of dark matter (DM) in our Universe.
However, the nature of DM remains one of the major unsolved problems in fundamental physics.
Primordial black holes (PBHs) produced in the radiation-dominated era of the early universe due to the collapse of large energy density fluctuations, as a promising candidate for dark matter, have attracted much attention~\cite{Hawking:1971ei,Carr:1974nx,Carr:1975qj,Khlopov:2008qy,Carr:2009jm,Carr:2016drx,Gao:2018pvq,Cai:2018rqf,Sasaki:2018dmp,Saito:2008jc,Cai:2018dig,Chen:2018rzo,Carr:2017jsz,Kannike:2017bxn,Kuhnel:2019xes,Kuhnel:2015vtw}.

Two neighboring PBHs can form a binary in the early Universe and coalesce within the age of the Universe.
The merge rate of PBH binaries was first estimated through the three-body interaction
for the case where all PBHs have the same mass~\cite{Nakamura:1997sm,Ioka:1998nz}.
In the PBH binary formation scenario,
the gravitational wave event GW150914 detected by LIGO~\cite{Abbott:2016blz} and
the merger rate estimated by the LIGO-Virgo Collaboration can be explained by
the coalescence of PBH binaries
if PBHs have the mass about $30 \Msun$ and constitute a tiny fraction of DM~\cite{Sasaki:2016jop}.
The binary formation was extended to account for an arbitrary PBH mass function
based on the three-body approximation~\cite{Raidal:2017mfl}
or to account for the torque from the surrounding PBHs as well as standard large-scale adiabatic perturbations
assuming a monochromatic mass function~\cite{Ali-Haimoud:2017rtz}.
The mechanism has recently been developed for a general mass function
by taking into account the torque from the surrounding PBHs~\cite{Kocsis:2017yty,Chen:2018czv,Raidal:2018bbj,Liu:2018ess}.

However, these studies ignore the possibility that a PBH binary merges into a new black hole
which together with another PBH form a new PBH binary.
Such a second-merge event can in principle be detected by LIGO-Virgo at the present time.
In this paper, we develop an analytic formalism to work out the merger rate density of PBH binaries with a general mass function, by taking into account the merger history of PBHs.

The paper is organized as follows.
In the next section, we summarize the basic equation for the primordial input parameters of PBHs
and revisit the merger rate for a monochromatic mass function as the first-merger process.
In Sec.~\ref{Formalism}, we develop a formalism to calculate the merger rate density of PBH binaries with a general mass function,
by taking into account the merger history of PBHs.
In Sec.~\ref{Examples}, we consider three specific examples, monochromatic, power-law mass and log-normal functions,
to investigate the effects of the merger history on the merger rate density of PBH binaries.
The final section is devoted to conclusions.

In this paper, we use units of $c = G = 1$.
Whenever relevant, we adopt the values of cosmological parameters consistent with the Planck measurements~\cite{Ade:2015xua}.
The scale factor is normalized to unity at the present time.

\section{Single-merger events}
\label{PBH merger}
Let us start with deriving the basic equation of the merger rate of PBH binaries.
It could be easily checked that the gravitational attraction between two approximately isolated PBHs dominates their dynamics if their average mass is bigger than  the background mass contained in a comoving sphere whose radius equals to their conformal distance. Considering the different scaling with time of the two competing effects (their gravitational attraction versus the expansion of the Universe) in the equation of motion for their separation~\cite{Ali-Haimoud:2017rtz}.
Following Ref.~\cite{Sasaki:2016jop}, in this section, we assume that all PBHs have the same mass, $M$,
and PBH binaries decouple from the expansion of the Universe during radiation domination provided that their comoving separation, $x$,  approximately satisfies
\begin{equation}
x <  x_{\rm max} \equiv (f_{\rm pbh}/n_{\rm pbh})^{1/3}=\(M/\rho_{\rm dm}\)^{1/3},
\label{eqn:condition_dec}
\end{equation}
where $f_{\rm pbh}$ is the fraction of PBHs in DM,
$n_{\rm pbh}$ denotes the comoving average number density of PBHs and $\rho_{\rm dm}$ denotes the present energy density of DM. The redshift $z_{\rm dec}$ at which the binary decoupling occurs is given by
\begin{equation}
1+z_{\rm dec}=(1+z_{\rm eq}) \left({x_{\rm max}}/{x}\right)^3,
\end{equation}
where $z_{\rm eq}\simeq 3400$ is the redshift at matter-radiation equality,
assuming negligible initial peculiar velocities here and throughout.
Therefore, given PBH mass $M$ and the initial comoving distance of PBHs $x$,
the decoupling time is determined by PBHs. In this work,
we assume that accretion and evaporation are negligible before the epoch of binary formation.
When two PBHs come closer, the nearest PBH exert torque on the bound system.
As a result, the two PBHs avoid a head-on collision and form a highly eccentric binary.
The major and minor axes are given by (denoted by $a$ and $b$, respectively)
\m
\label{a}
a = A \frac{x}{1+z_{\rm dec}} = A \frac{\rho_{\rm dm} x^4}{(1+z_{\rm eq})M},
\n
\m
b = B \(\frac{x}{y}\)^3 a,
\n
where $y$ is the comoving distance to the third PBH,
$A$ and $B$ are numerical factors of ${\cal O}(1)$.
A detailed investigation of the dynamics of the binary formation suggests $A=0.4$ and $B=0.8$~\cite{Ioka:1998nz}.
To be exact, in the following calculation, we adopt $A=0.4$ and $B=0.8$.
The dimensionless angular momentum of PBH binaries is given by
\m
\label{j1}
j \equiv  \sqrt{1-e^2} = B \(\frac{x}{y}\)^3,
\n
where $e$ is the eccentricity of the binary at the formation time.
Once two PBHs form a binary,
they gradually shrink through the emission of gravitational radiation and eventually merge at the time $\tau$ after its formation,
which can be estimated as~\cite{Peters:1964zz}
\begin{equation}
\label{tmerge}
\tau \simeq \frac{3 { a}^4 j^7}{170 M^3}.
\end{equation}

\begin{figure}[htbp!]
\centering
\includegraphics[width = 0.48\textwidth]{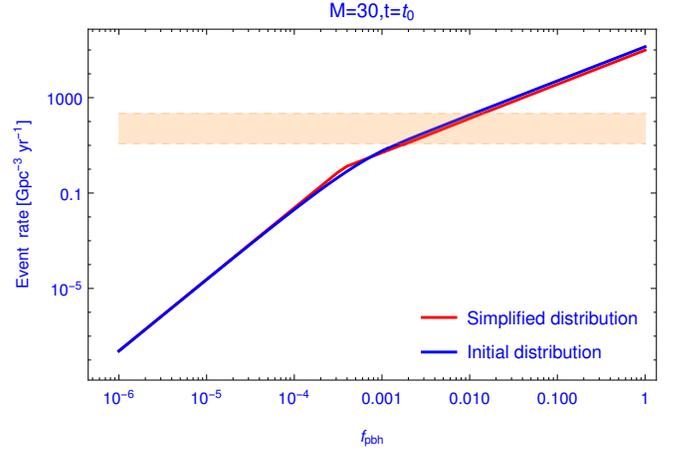}
\caption{\label{fig:distribution}
Event rate of single-mergers of PBH binaries with the mass of $30 \Msun$ as
a function of the PBH abundance.
The blue line corresponds to the case of the distribution~\eqref{distribution 1}
and the red line corresponds to the case of the distribution~\eqref{distribution 2}.
The merger rate $R=12 - 213$ Gpc$^{-3}$ yr$^{-1}$ inferred by the LIGO and Virgo Collaboration is shown as the shaded region colored orange~\cite{Abbott:2017vtc}. }
\end{figure}

To calculate the merger rate of PBH binaries, we have to know the spatial distribution of PBHs.
Assuming that the spatial distribution of PBHs is random one,
the probability that the comoving distances, $x$ and $y$, are in the intervals $(x,x+dx)$ and $(y,y+dy)$  is given by
\m \label{distribution 1}
dP=\frac{4\pi x^2dx}{n_{\rm pbh}^{-1}} \frac{4\pi y^2dy}{n_{\rm pbh}^{-1}}
\exp \left( -\frac{4\pi y^3}{3n_{\rm pbh}^{-1}} \right) \Theta (y-x).
\n
To deal with this probability distribution, we can rewrite Eq.~\eqref{distribution 1} as follows
\m \label{distribution 2}
dP=\frac{4\pi x^2dx}{n_{\rm pbh}^{-1}} \frac{4\pi y^2dy}{n_{\rm pbh}^{-1}} \Theta (y-x)\Theta (y_{\rm max}-y),
\n
where $y_{\rm max}={\left( 4\pi n_{\rm pbh} /3 \right)}^{-1/3}$, which is adopted in~\cite{Nakamura:1997sm}.
In Fig.~\ref{fig:distribution} we show the merger rate estimated by using the initial distribution~\eqref{distribution 1}
and the simplified distribution~\eqref{distribution 2},
which indicates that the difference between the two cases is insignificant
compared to the uncertainty of the merger rate estimated
by the LIGO-Virgo Collaboration.

\begin{figure}[htbp!]
\centering
\includegraphics[width = 0.48\textwidth]{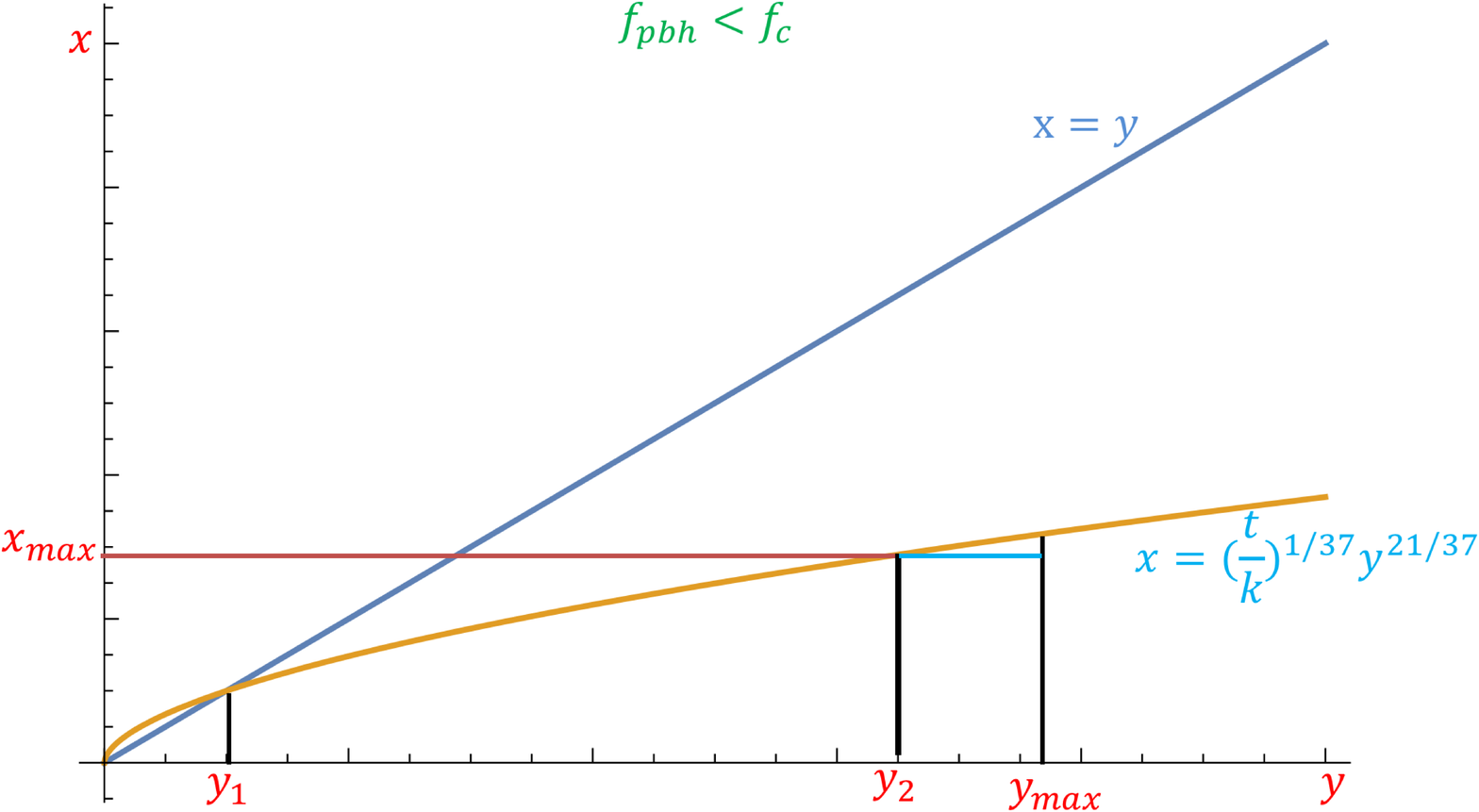}
\includegraphics[width = 0.48\textwidth]{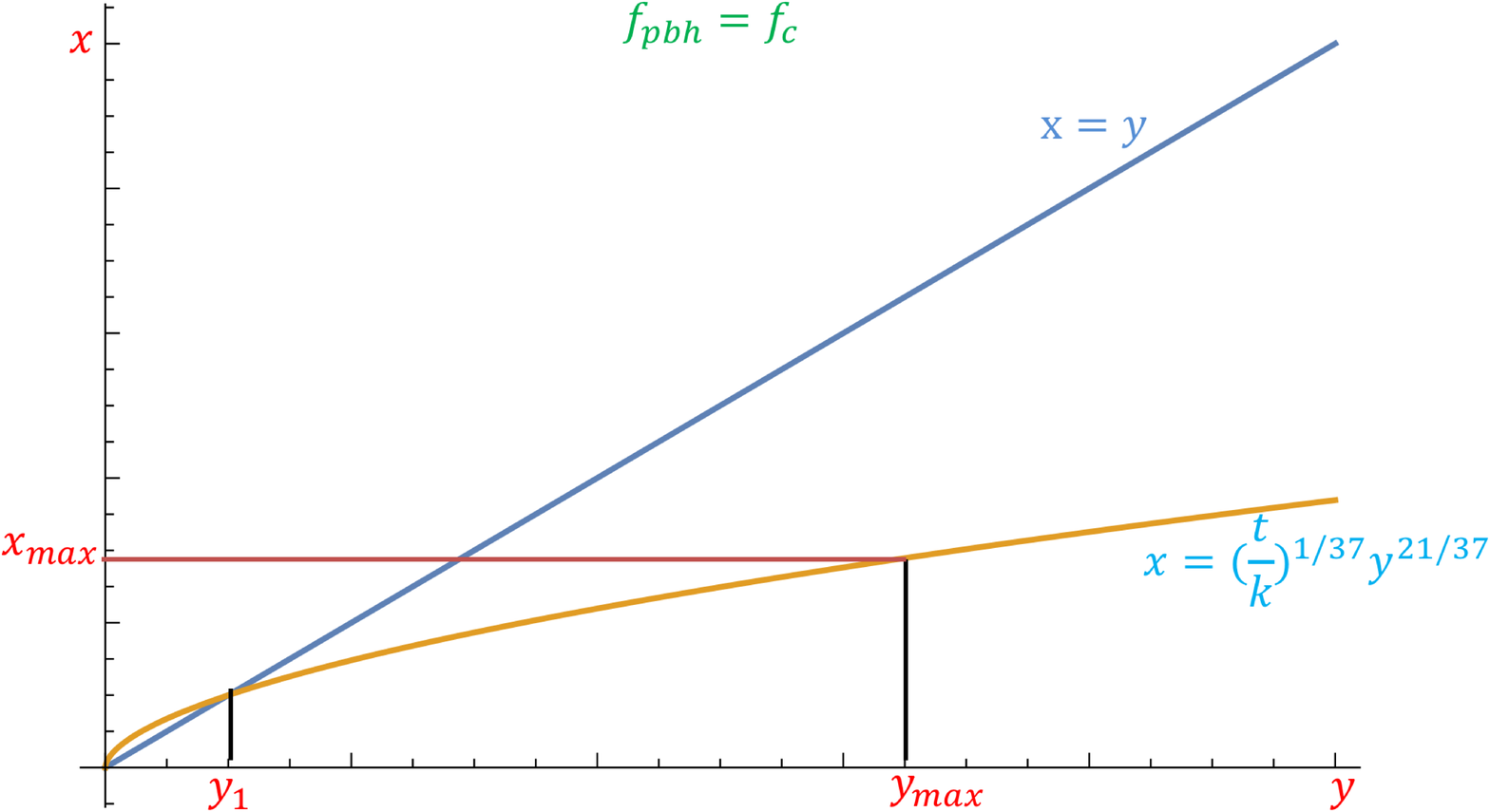}
\includegraphics[width = 0.48\textwidth]{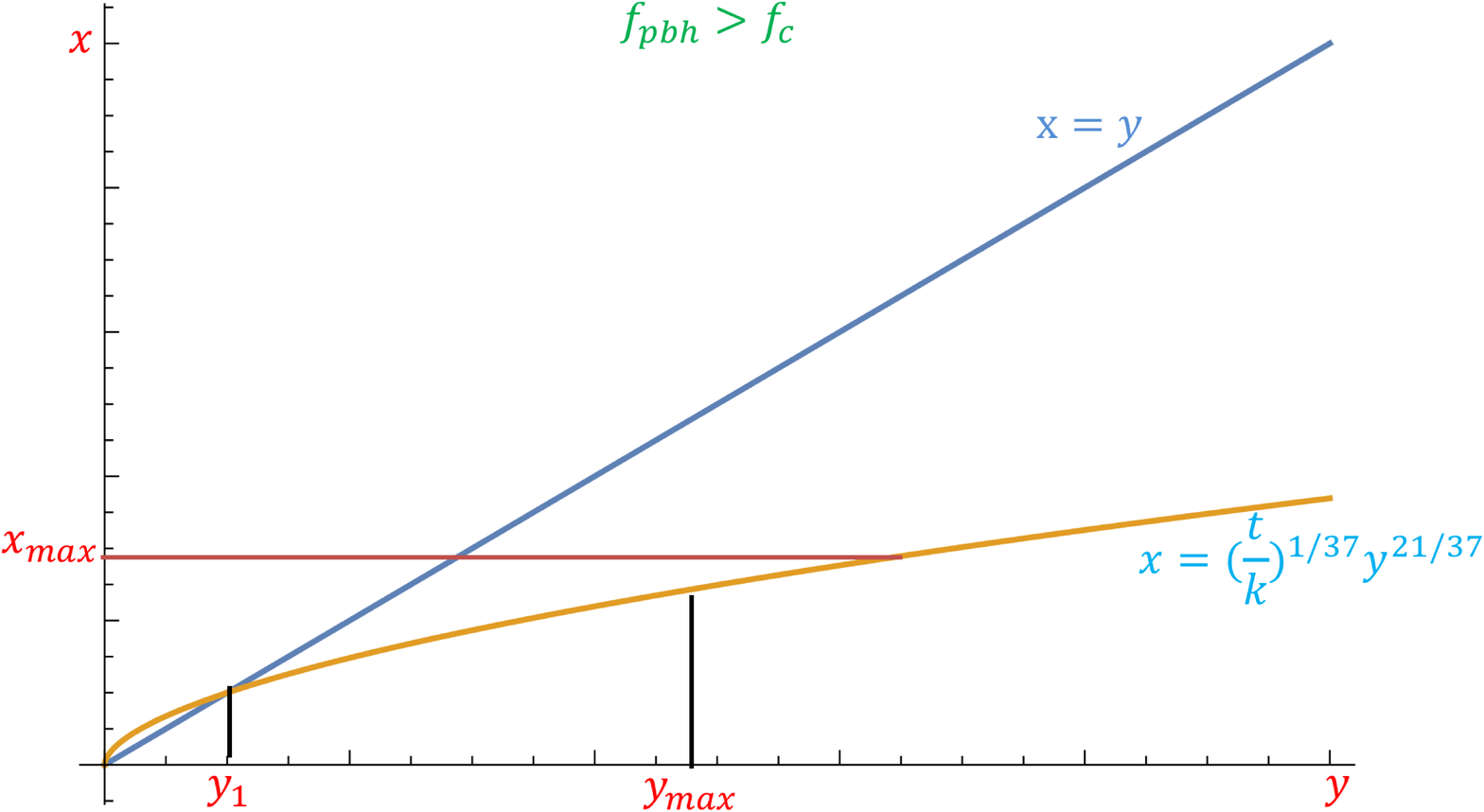}
\caption{\label{fig:illustration}
Schematic illustration on calculating $G(t)$}
\end{figure}

The fraction of PBHs which have merged before the time $t$ is given by
\m
G(t)=\int dxdy \frac{d P}{dxdy} \Theta (t-\tau(x,y))\Theta(x_{\rm max}-x).
\n
In Fig.~\ref{fig:illustration} is a schematic illustration on calculating $G(t)$.

\begin{figure}[htbp!]
\centering
\includegraphics[width = 0.48\textwidth]{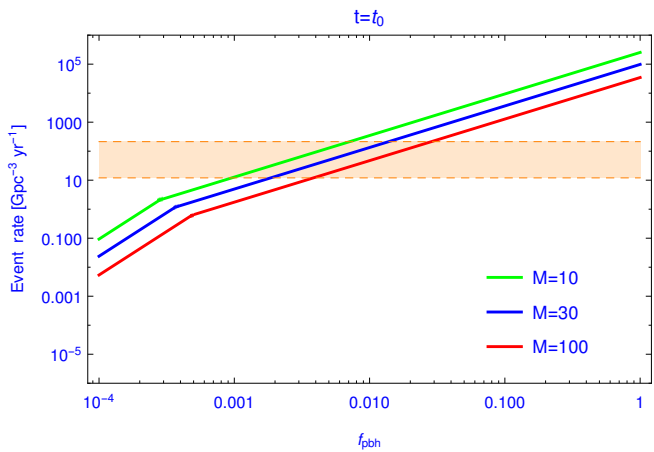}
\caption{\label{fig:merger}
Event rate of single-mergers of PBH binaries with the mass $M=10$ (green), $30$ (blue) and $100$ (red) in $\Msun$ at the present time as a function of the PBH abundance.
The merger rate $R=12 - 213$ Gpc$^{-3}$ yr$^{-1}$ inferred by the LIGO and Virgo Collaboration is shown as the shaded region colored orange~\cite{Abbott:2017vtc}.}
\end{figure}

From Eqs.~\eqref{a}, \eqref{j1} and \eqref{tmerge}, we can get
\m
x=(\frac{t}{k})^{1 \over 37} y^{21 \over 37},
\n
where
\m
k=\frac{3}{170} \frac{1}{M^3} (\frac{\rho_{\rm dm}}{(1+z_{\rm eq})M})^{4} A^{4} B^{7}.
\n
When $f=f_{c}$, there is 
\m
\label{xymax}
x_{\rm max}=(\frac{t}{k})^{1 \over 37} y_{\rm max}^{21 \over 37}
\n
By solving Eq.~\eqref{xymax}, we can get
\m
f_{c}&=&(\frac{4\pi}{3})^{-1}(\frac{t}{k})^{1\over 7} \(M \over \rho_{\rm dm}\)^{-{16 \over 21}}
\nonumber \\
&\approx& 1.63 \times 10^{-4} \({M \over \Msun}\)^{{5 \over 21}} \({t\over t_0}\)^{{1\over 7}}
\n
By solving
\m
y_1=x, ~~~~x=(\frac{t}{k})^{1 \over 37} y_1^{21 \over 37},
\n
we arrive
\m
y_1=(\frac{t}{k})^{1 \over 16}.
\n
By solving
\m
x_{\rm max}=(\frac{t}{k})^{1 \over 37} y_{2}^{21 \over 37}
\n
we arrive
\m
y_{2}=(\frac{k}{t})^{1 \over 21} x_{\rm max}^{37 \over 21}
\n
For $f>f_c$, $G(t)$ is given by
\m
G(t)&=& \int _{0}^{y_1} \int _{0}^{y} 4\pi x^2 n_{\rm pbh} 4\pi y^2 n_{\rm pbh} dxdy
\nonumber \\
&+&\int _{y_1}^{y_{max}} \int _{0}^{(\frac{t}{k})^{1/37} y^{21/37}} 4\pi x^2 n_{\rm pbh} 4\pi y^2 n_{\rm pbh} dxdy
\nonumber \\
&=& \frac{8 \pi^2}{261} (37 y_{max}^{174/37} (\frac{t}{k})^{3/37}-8 (\frac{t}{k})^{3/8}) n_{\rm pbh}^{2}
\nonumber \\
&\approx&2.85\times 10^{-3} \({ M \over\Msun}\)^{{5 \over 37}} \({t\over t_0}\)^{{3\over 37}} \fpbh^{16 \over 37}
\nonumber \\
&-& 1.86\times 10^{-12}\({ M \over\Msun}\)^{{5 \over 8}} \({t\over t_0}\)^{{3\over 8}} \fpbh^2
\nonumber \\
&\approx& 2.85\times 10^{-3} \({ M \over\Msun}\)^{{5 \over 37}} \({t\over t_0}\)^{{3\over 37}} \fpbh^{16 \over 37}
\n
For $f<f_c$, $G(t)$ is given by
\m
G(t)&=& \int _{0}^{y_1} \int _{0}^{y} 4\pi x^2 n_{\rm pbh} 4\pi y^2 n_{\rm pbh} dxdy
\nonumber \\
&+& \int _{y_1}^{y_2} \int _{0}^{(\frac{t}{k})^{1 \over 37} y_{2}^{21 \over 37}} 4\pi x^2 n_{\rm pbh} 4\pi y^2 n_{\rm pbh} dxdy
\nonumber \\
&+& \int _{y_2}^{y_{\rm max}} \int _{0}^{x_{\rm max}} 4\pi x^2 n_{\rm pbh} 4\pi y^2 n_{\rm pbh} dxdy
\nonumber \\
&=&-\frac{8 \pi^2}{261}(8 (\frac{t}{k})^{3\over 8}+(\frac{t}{k})^{3\over 37}y_2^{63\over 37}(-58y_{\rm max}^3+21y_2^3)) n_{\rm pbh}^{2}
\nonumber \\
&\approx& \fpbh(4.19-1.18\times10^{-8} \({ M \over\Msun}\)^{{5 \over 8}} \({t\over t_0}\)^{{3\over 8}} \fpbh
\nonumber \\
&-& 9.32\times 10^{3} \({ M \over\Msun}\)^{-{5 \over 21}} \({t\over t_0}\)^{-{1\over 7}} \fpbh)
\nonumber \\
&\approx& \fpbh(4.19-9.32\times 10^{3} \({ M \over\Msun}\)^{-{5 \over 21}} \({t\over t_0}\)^{-{1\over 7}} \fpbh)
\nonumber \\
\n
Therefore, the merger rate of PBH binaries per unit volume per unit time (at the time $t$) can be easily obtained by
\m \label{merger rate 1}
R\(t\)=\frac{1}{2} n_{\rm pbh}\lim _{ dt\rightarrow 0 }{ \frac{G\(t+dt\)-G\(t\)}{dt}},
\n
where the factor $1/2$ accounts for that each merger event involves two PBHs.
From Eq.~\eqref{merger rate 1}, the final result is given by
\begin{equation} \label{R1}
\begin{split}
&R\(t\) \approx  \\
 &\begin{cases}
   1.61 \times 10^{12} \({M \over \Msun}\)^{-{26 \over 21}} \({t\over t_0}\)^{-{8\over 7}} \fpbh^3, ~~~~{\rm for}~\fpbh<f_{c},\\
    1.86 \times 10^{6} \({M \over \Msun}\)^{-{32 \over 37}} \({t\over t_0}\)^{-{34\over 37}} \fpbh^{53 \over 37},~~~~{\rm for}~\fpbh \ge f_{c},
  \end{cases}
  \end{split}
\end{equation}
which can be interpreted as the merger rate in Gpc$^{-3}$ yr$^{-1}$.
We show the single-merger rate of PBH binaries as a function of the PBH abundance in Fig.~\ref{fig:merger}.
For $\fpbh>f_{c}$ it scales as $\fpbh^{53/37}$ and for $\fpbh<f_{c}$ it scales as $\fpbh^3$.

Now we have to emphasize what is the difference between our formalism
and the one developed in Refs~\cite{Sasaki:2016jop}.
For the single-merger case, the merger rate of PBH binaries
is usually calculated by converting the probability distribution function $P$ of $x$ and $y$
into the one of $a$ and $e$.
However, for the multiple-merger case, the probability distribution is a function of more variables than $x$ and $y$.
It becomes hard to convert the probability distribution function of $(x,y,z,...)$ into the one of $a$ and $e$.
Therefore, the known formalism does not work in the multiple-merger case.
To get the merger rate of PBH binaries, we directly deal with the probability distribution
in the $x-y$ plane to find which PBHs have been merged.
It becomes easy to extend our formalism to the second and third merger events.
In this section, to warm up we consider the merger rate of PBH binaries in the single-merger case.
In the next section, we shall extend the formalism to the second- and third-merger cases.

\section{Multiple-merger events}
\label{Formalism}
So far, several gravitational wave events from black hole binary mergers have been detected by the LIGO-Virgo collaboration,
such as GW150914 ($36_{-4}^{+5}M_\odot$, $29_{-4}^{+4}M_\odot$)~\cite{Abbott:2016blz}, GW151226 ($14.2_{-3.7}^{+8.3}M_\odot$, $7.5_{-2.3}^{+2.3}M_\odot$)~\cite{Abbott:2016nmj}, GW170104 ($31.2_{-6.0}^{+8.4}M_\odot$, $19.4_{-5.9}^{+5.3}M_\odot$)~\cite{Abbott:2017vtc}, GW170608 ($12_{-2}^{+7}M_\odot$, $7_{-2}^{+2}M_\odot$)~\cite{Abbott:2017gyy} and GW170814 ($30.5_{-3.0}^{+5.7}M_\odot$, $25.3_{-4.2}^{+2.8}M_\odot$)~\cite{Abbott:2017oio}.
These events detected by LIGO-Virgo suggest that the black holes should have an extended mass function.
In this section, we calculate the merger rate distribution for PBH binaries with a general mass function by taking into account the effect of merger history on the merger rate density of PBH binaries.

First of all, we consider the condition that two neighboring PBHs with the masses $m_i$ and $m_j$
decouple from the expansion of the Universe and form a bound system.
Their comoving separation, $x$, approximately satisfies
\m
\label{xmax}
x <  x_{\rm max} =\(\frac{m_b}{2\rho_{\rm dm}}\)^{1/3},
\n
where $m_b=m_i+m_j$ is the total mass of the PBH binary.
When two PBHs come closer, the nearest PBH with the mass $m_l$, exert torque on the bound system.
As a result, the two PBHs avoid a head-on collision and form a highly eccentric binary.
The major axis $a$ of the binary orbit and the dimensionless angular momentum are given by
\m
a \approx A \frac{2\rho_{\rm dm} x^4}{(1+z_{\rm eq})m_b},
\n
\m \label{j}
j \approx B \frac{2 m_l}{m_b} \left(\frac{x}{y}\right)^3,
\n
where $y$ is the comoving distance to the third PBH with the mass $m_l$.
Once two PBHs form a binary,
they gradually shrink through the emission of gravitational radiation and eventually merge at the time $\tau$ after its formation,
which can be estimated as~\cite{Peters:1964zz}
\m   \label{coalescence time}
  \tau = \frac{3}{85} \frac{a^4}{m_i m_j m_b} j^7.
\n

The two neighboring PBHs with the masses $m_i$ and $m_j$ merge into a bigger black hole.
The mass is given by
\m
M_2=m_b-E_{\rm GW} \approx \gamma m_b,
\n
where $E_{\rm GW}$ is the energy of gravitational wave and $\gamma$ is a factor of ${\cal O}(1)$.
In the monochromatic case, $\gamma =0.95$ is adopted in~\cite{Bringmann:2018mxj}.
For simplicity, in this paper, we take $\gamma=1$, which means we assume that the energy of gravitational wave is zero.

In this paper, the probability distribution function of PBHs $P(m)$ is normalized to be
\m
\int dm P(m) = 1.
\n
Therefore, the abundance of PBHs in the mass interval $(m, m+dm)$ can be easily obtained by
\m
f P(m)dm, 
\n
where
$f$ is a fraction of PBHs in non-relativistic matter including DM and baryons.
The fraction of PBHs in DM $f_{\rm{pbh}}$ is given by $f_{\rm{pbh}}\equiv \Omega_{\rm{pbh}}/\Omega_{\rm{dm}} \approx f/0.85$.
At the present time, the average number density of PBHs in the mass interval $(m, m+dm)$ is given by
\m
n\(m\)dm= \frac{fP(m)dm \rho_{\rm{m}}}{m}= \frac{f_{\rm{pbh}} P(m)dm \rho_{\rm{dm}}}{m},
\n
where $\rho_{\rm{m}}$ is the total energy density of matter and the present total average number density of PBHs, $n_{T}$, is obtained by
\m
n_T\equiv f_{\rm{pbh}}\rho_{\rm{dm}}\int dm {P(m)\over m}.
\n
For simplicity, here we define $m_{\rm pbh}$ as
\m
 \label{mpbh}
\frac{1}{m_{\rm{pbh}}}=\int dm {P(m)\over m} .
\n
We define $F(m)$ as
\m
 \label{F(m)}
F\(m\) \equiv \frac{n\(m\)}{n_{T}}=P\(m\) \frac{m_{\rm{pbh}}}{m},
\n
which is the fraction of the present average number density of PBHs with the mass $m$ in the present total average number density of PBHs.

The result in~\cite{Ali-Haimoud:2017rtz} indicates that in the case of $\fpbh < f_{c}$,
the effects of the linear density perturbations on the merger rate of PBH binaries is significant.
Here, we only consider the the case of $\fpbh >f_{c}$ which is shown to be relevant to
the LIGO observations~\cite{Sasaki:2016jop}. In other words, we ignore the bound~\eqref{xmax}.

The only essential ingredient that we need is the spatial distribution of PBHs. We firstly consider the spatial distribution of two PBHs. The probability distribution of the comoving separation $x$ between two nearest PBHs with the masses $(m_i, m_i+dm_i)$ and $(m_j, m_j+dm_j)$ and without other PBHs
in the comoving volume of $4\pi x^3/3$ is given by
\m
 \label{distribution}
&d\hat{P}&\(m_i,m_j,x\)=F\( m_i\)dm_i 4\pi x^2 dx  n\(m_j\)dm_j
\nonumber \\
&\times& e^{- {4\pi \over 3}x^3 n(m_j) dm} \prod _{ m\neq m_j }^{ }{  e^{- {4\pi \over 3}x^3 n(m) dm}}
\nonumber \\
&=&F\( m_i\)dm_i 4\pi x^2dx n\(m_j\)dm_j e^{- \int dm {4\pi \over 3}x^3 n(m)}
\nonumber \\
&=&F\( m_i\)dm_i F\( m_j\)dm_j 4\pi x^2 n_{T}dx
e^{- {4\pi \over 3}x^3 n_{T}}.
 \nonumber  \\
\n
Clearly, in the non-monochromatic case, to calculate the merger rate in the first-merger process,
the differential probability distribution is given by
\m
 \label{distribution 5}
&d{P_{1}}&\(m_i,m_j,m_l,x,y\)=F\( m_i\)dm_i F\( m_j\)dm_j  F\( m_l\)dm_l
\nonumber \\
&\times& 4\pi x^2 n_{T}dx 4\pi y^2 n_{T}dy
e^{- {4\pi \over 3}y^3 n_{T}} \Theta(y-x),
\n
where $x$ is the comoving separation between two nearest PBHs with the masses $m_i$ and $m_j$ and
$y$ is the comoving distance to the third PBH with the mass $m_l$ which provides the angular momentum for the bound system.
The fraction of PBHs that have merged before the time $t$ is given by
\m
&{G_1}&(t,m_i,m_j,m_l)
\nonumber \\
&=&\int dxdy \frac{d P_{1}\(m_i,m_j,m_l,x,y\)}{dx dy dm_i dm_j dm_l} \Theta (t-\tau(x,y)).
\nonumber \\
\n
So, we can arrive
\m
 \label{G1bar}
&G_1&(t,m_i,m_j,m_l) =F\( m_i\) F\( m_j\)  F\( m_l\)
\nonumber \\
& \times& 1.34\times 10^{-2} \({ \Msun}\)^{-{5 \over 37}} \({t\over t_0}\)^{{3\over 37}} \({ m_i m_j}\)^{{3 \over 37}}
\nonumber \\
 & \times& \({ m_l}\)^{-{21 \over 37}} \({ m_{\rm pbh}}\)^{-{16 \over 37}} \({ m_i+m_j}\)^{{36 \over 37}}
 \fpbh^{16 \over 37}.
\n
$\RR_1(t,m_i,m_j,m_l)$ is given by
\m
 \label{Rbar 1}
&\RR_1&(t,m_i,m_j,m_l)=\frac{1}{2} n_{\rm T}
\nonumber \\
&\times& \lim _{ dt\rightarrow 0 }{ \frac{G_1\(t+dt,m_i,m_j,m_l\)-G_1\(t,m_i,m_j,m_l\)}{dt}}.
\nonumber \\
\n
where the factor $1/2$ accounts for that each merger event involves two PBHs.
From Eq.~\eqref{Rbar 1}, one has
\m
\label{R1bar}
&\RR_1&(t,m_i,m_j,m_l) =F\( m_i\) F\( m_j\)  F\( m_l\)
\nonumber \\
& \times& 1.32\times 10^{6} \({ \Msun}\)^{{32 \over 37}} \({t\over t_0}\)^{-{34\over 37}} \({ m_i m_j}\)^{{3 \over 37}}
\nonumber \\
 & \times& \({ m_l}\)^{-{21 \over 37}} \({ m_{\rm pbh}}\)^{-{53 \over 37}} \({ m_i+m_j}\)^{{36 \over 37}}
 \fpbh^{53 \over 37}.
\n
The merger rate density of PBH binaries with the masses $m_i$ and $m_j$ in the first-merger process is
\m
\label{R1}
\RR_1(t,m_i,m_j)=\int dm_l \RR_1(t,m_i,m_j,m_l) .
\n

Let us estimate the merger rate density in the second-merger process.
In the first-merger process, two neighboring PBHs decouple from the expansion of the Universe
and then merge into a new black hole with the mass $m_i+m_j$.
In the second-merger process, the new black hole and the nearest PBH with mass $m_k$ form a new binary.
The merge event of the new binary is detected by LIGO-Virgo at the time $t$.
Statistically, the second coalescence time is larger than the first one, therefore, we can ignore the first coalescence time. The differential probability distribution is given by
\m
 \label{distribution 7}
&d{P_{2}}\(m_i,m_j,m_k,m_l,x,y,z\)
\nonumber \\
&=F\( m_i\)dm_i F\( m_j\)dm_j  F\( m_k\)dm_k F\( m_l\)dm_l
\nonumber \\
& 4\pi x^2 n_{T}dx 4\pi y^2 n_{T}dy 4\pi z^2 n_{T}dz
e^{- {4\pi \over 3}z^3 n_{T}} \Theta(y-x) \Theta(z-y).
\nonumber \\
\n
So, the fraction of PBHs that have merged in the second-merger process is given by
\m
&&{G_2}(t,m_i,m_j,m_k,m_l)
\nonumber \\
&&=\int dxdydz \frac{d P_{2}\(m_i,m_j,m_k,m_l,x,y\)}{dx dy dz dm_i dm_j dm_k dm_l} \Theta (t-\tau(y,z)).
\nonumber \\
\n
Then, we can arrive
\m
&G_2&(t,m_i,m_j,m_k,m_l) =F\( m_i\) F\( m_j\) F\( m_k\)  F\( m_l\)
\nonumber \\
& \times& 1.21\times 10^{-4} \({ \Msun}\)^{-{10 \over 37}} \({t\over t_0}\)^{{6\over 37}} \({ m_i+ m_j}\)^{{6 \over 37}}  \({ m_k}\)^{{6 \over 37}}
\nonumber \\
 & \times& \({ m_l}\)^{-{42 \over 37}} \({ m_{\rm pbh}}\)^{-{32 \over 37}} \({ m_i+m_j+m_k}\)^{{72 \over 37}} \fpbh^{32 \over 37}.
\n
$\RR_2(t,m_i,m_j,m_l)$ is given by
\m
 \label{Rbar 2}
&\RR_2&(t,m_i,m_k,m_j,m_l)=\frac{1}{3} n_{\rm T}
\nonumber \\
&\times& \lim _{ dt\rightarrow 0 }{ \frac{G_2\(t+dt,m_i,m_j,m_k,m_l\)-G_2\(t,m_i,m_j,m_k,m_l\)}{dt}},
\nonumber \\
\n
where the factor $1/3$ accounts for that each merger event in second-merger process involves three PBHs.
From Eq.~\eqref{Rbar 2}, the final result is given by
\m
&\RR_2&(t,m_i,m_j,m_k,m_l) =F\( m_i\) F\( m_j\) F\( m_k\)  F\( m_l\)
\nonumber \\
& \times& 1.59\times 10^{4} \({ \Msun}\)^{{27 \over 37}} \({t\over t_0}\)^{-{31\over 37}} \({ m_i+ m_j}\)^{{6 \over 37}}  \({ m_k}\)^{{6 \over 37}}
\nonumber \\
 & \times& \({ m_l}\)^{-{42 \over 37}} \({ m_{\rm pbh}}\)^{-{69 \over 37}} \({ m_i+m_j+m_k}\)^{{72 \over 37}} \fpbh^{69 \over 37}.
\n

The merger rate density of PBH binaries with the masses $m_i$ and $m_j$ in the second-merger process is given by
\m
&\RR_2&(t,m_i,m_j)=\frac{1}{2}\int dm_ldm_e \RR_2(t,m_i-m_e,m_e,m_j,m_l)
\nonumber \\
&+& \frac{1}{2}\int dm_ldm_e \RR_2(t,m_j-m_e,m_e,m_i,m_l) .
\n

Similarly, $G_3(t,m_i,m_j,m_k,m_f,m_l)$ and $\RR_3(t,m_i,m_j,m_k,m_f,m_l)$ are given by
\m
&G_3&(t,m_i,m_j,m_k,m_f,m_l)
 \nonumber \\
&=&F\( m_i\) F\( m_j\) F\( m_k\) F\( m_f\)  F\( m_l\)
\nonumber \\
& \times& 8.88\times 10^{-7} \({ \Msun}\)^{-{15 \over 37}} \({t\over t_0}\)^{{9\over 37}} \({ m_i+ m_j+m_k}\)^{{9 \over 37}}  \({ m_f}\)^{{9 \over 37}}
\nonumber \\
 & \times& \({ m_l}\)^{-{63 \over 37}} \({ m_{\rm pbh}}\)^{-{48 \over 37}} \({ m_i+m_j+m_k+m_f}\)^{{108 \over 37}} \fpbh^{48 \over 37}.
 \nonumber \\
\n
\m
\label{R3bar}
&\RR_3&(t,m_i,m_j,m_k,m_f,m_l)
 \nonumber \\
&=&F\( m_i\) F\( m_j\) F\( m_k\) F\( m_f\)  F\( m_l\)
\nonumber \\
& \times& 1.31\times 10^{2} \({ \Msun}\)^{{22 \over 37}} \({t\over t_0}\)^{-{28\over 37}} \({ m_i+ m_j+m_k}\)^{{9 \over 37}}  \({ m_f}\)^{{9 \over 37}}
\nonumber \\
 & \times& \({ m_l}\)^{-{63 \over 37}} \({ m_{\rm pbh}}\)^{-{85 \over 37}} \({ m_i+m_j+m_k+m_f}\)^{{108 \over 37}} \fpbh^{85 \over 37}.
 \nonumber \\
\n
The merger rate density of PBH binaries with the masses $m_i$ and $m_j$ in the third-merger process is given by
\m
\label{R3}
&\RR_3&(t,m_i,m_j)
 \nonumber \\
 &=& \frac{1}{2} \int dm_ldm_edm_f \RR_3 (t,m_i-m_e-m_f,m_e,m_f,m_j,m_l)
 \nonumber \\
 &+&  \frac{1}{2} \int dm_ldm_edm_f \RR_3 (t,m_j-m_e-m_f,m_e,m_f,m_i,m_l) .
  \nonumber \\
\n
The total merger rate density of PBH binaries with the masses $m_i$ and $m_j$ detected by LIGO-Virgo is given by
\m
\RR(t,m_i,m_j) = \sum_{n=1} \RR_n(t,m_i,m_j) \,.
\n
In the single-merger case, we have $\alpha = -(m_i+m_j)^2\p^2 \ln \RR (t,m_i,m_j)/\p m_i \p m_j=36/37$
which is independent of the PBH mass function.
It is consistent with the result obtained in~\cite{Kocsis:2017yty}.
However, by taking account into the merger history of PBHs,
$\alpha$ depends on the PBH mass function,
which could help us reconstruct the mass function of PBHs.

\begin{figure}[htbp!]
\centering
\includegraphics[width = 0.48\textwidth]{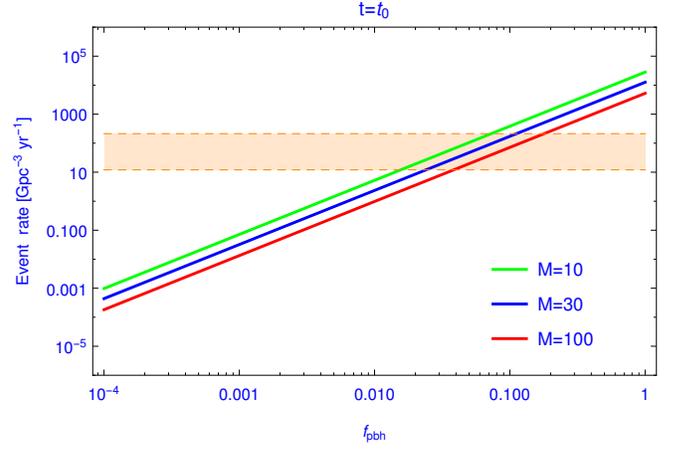}
\caption{\label{fig:merger2}
Event rate of second-merger of PBH binaries with the mass $M = 10$ (green), $30$ (blue) and $100$ (red) in $\Msun$
at the present time as a function of the PBH abundance.
The merger rate $R=12 - 213$ Gpc$^{-3}$ yr$^{-1}$ inferred by the LIGO and Virgo Collaboration is shown as the shaded region colored orange~\cite{Abbott:2017vtc}.}
\end{figure}

\section{Applications}
\label{Examples}
The total fraction of PBH binaries that have merged before the time $t$ in single-merger events is given by
\m
\label{G1}
G_1 (t) \equiv \int \int G_1(t,m_i,m_j,m_l) dm_idm_j dm_l.
\n
The merger rate of PBH binaries in single-merger events at the time $t$ is given by
\m
\label{R1G1}
R_1 (t) \equiv \frac{1}{2} n_{\rm T} \frac{dG_1(t)}{dt} \equiv \int\int \RR_1(t,m_i,m_j) dm_idm_j.
  \nonumber \\
\n
$G_N(t)$ is the total fraction of PBH binaries that have merged before the time $t$ in $N$-th merger process and $R_N(t)$ is merger rate of PBH binaries at time $t$ in $N$-th merger process.

Let us consider three typical PBH mass functions: monochromatic, power-law and log-normal function.

\subsection{Monochromatic mass function}
\begin{figure}[htbp!]
\centering
\includegraphics[width = 0.48\textwidth]{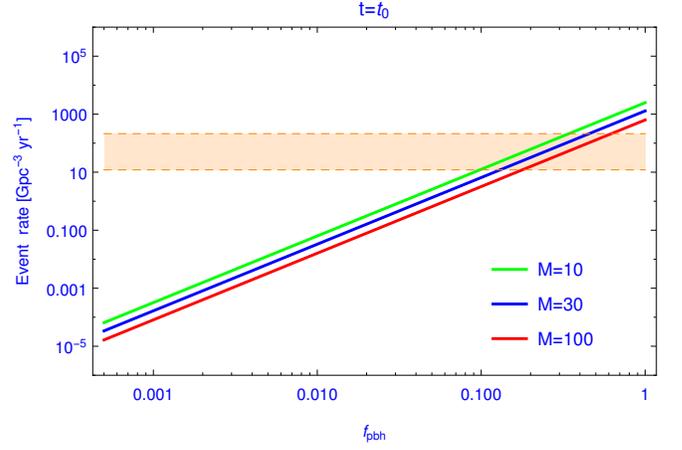}
\caption{\label{fig:merger3}
Event rate of third-merger of PBH binaries with the mass $M = 10$ (green), $30$ (blue) and $100$ (red) in $\Msun$
at the present time as a function of the PBH abundance.
The merger rate $R=12 - 213$ Gpc$^{-3}$ yr$^{-1}$ inferred by the LIGO and Virgo Collaboration is shown as the shaded region colored orange~\cite{Abbott:2017vtc}.}
\end{figure}

\begin{figure}[htbp!]
\centering
\includegraphics[width = 0.48\textwidth]{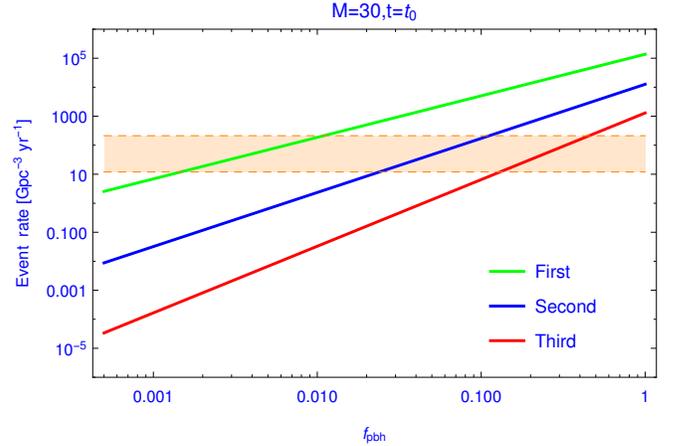}
\caption{\label{fig:mergertotal}
Event rate of first-merger (green), second-merger (blue) and third-merger (red) of PBH binaries with the mass $30\Msun$ at the present time as a function of the PBH abundance.
The merger rate $R=12 - 213$ Gpc$^{-3}$ yr$^{-1}$ inferred by the LIGO and Virgo Collaboration is shown as the shaded region colored orange~\cite{Abbott:2017vtc}.}
\end{figure}

In this subsection, we consider the following monochromatic mass function~\cite{Sasaki:2016jop,Bird:2016dcv,Nishikawa:2017chy}
\m
P(m)=\delta(m-M).
\n
In this case, we can rewrite~\eqref{mpbh} and~\eqref{F(m)} as
\m
\label{mpbh1}
m_{\rm{pbh}}=M,
\n
\m
\label{F(m)1}
F(m)=P(m)=\delta(m-M).
\n
From Eqs.~\eqref{G1bar}, \eqref{R1bar}, \eqref{R1}, \eqref{G1}, \eqref{R1G1}, \eqref{mpbh1} and~\eqref{F(m)1}, the total fraction of PBH binaries that have merged before the time $t$ and the merger rate of PBH binaries at the time $t$ in the first-merger process are given by
\m
G_1 \(t\) \approx 2.64 \times 10^{-2} \({M \over \Msun}\)^{{5 \over 37}} \({t\over t_0}\)^{{3\over 37}} \fpbh^{16 \over 37}
\n
\m
R_1\(t\) \approx 2.59 \times 10^{6} \({M \over \Msun}\)^{-{32 \over 37}} \({t\over t_0}\)^{-{34\over 37}} \fpbh^{53 \over 37}.
\n
which is consistent with~\eqref{merger rate 1}.
Similarly, the total fraction of PBH binaries that have merged before the time $t$ and the merger rate of PBH binaries at the time $t$ in the second-merger process are given by
\m
G_2 \(t\) \approx 1.15 \times 10^{-3} \({M \over \Msun}\)^{{10 \over 37}} \({t\over t_0}\)^{{6\over 37}} \fpbh^{32 \over 37}
\n
\m
 R_2\(t\) \approx 1.51 \times 10^{5} \({M \over \Msun}\)^{-{27 \over 37}} \({t\over t_0}\)^{-{31\over 37}} \fpbh^{69 \over 37}.
\n
In Fig.~\ref{fig:merger2}, we show the merger rate of PBH binaries in the second-merger process as a function of $\fpbh$,
which scales as $\fpbh^{69/37}$. The total fraction of PBH binaries that have merged before the time $t$ and the merger rate of PBH binaries at the time $t$ in the third-merger process are given by
\m
G_3 \(t\) \approx 6.64 \times 10^{-5} \({M \over \Msun}\)^{{15 \over 37}} \({t\over t_0}\)^{{9\over 37}} \fpbh^{48 \over 37}
\n
 \m
 R_3 \(t\)\approx 9.78 \times 10^{3} \({M \over \Msun}\)^{-{22 \over 37}} \({t\over t_0}\)^{-{28\over 37}} \fpbh^{85 \over 37}.
\n
In Fig.~\ref{fig:merger3}, we show the merger rate of PBH binaries in the third-merger process as a function of $\fpbh$,
which scales as $\fpbh^{69/37}$.
In the case of $M=30\Msun$ and $\fpbh=0.01$, we can find $R_1(t_0)=187$ Gpc$^{-3}$ yr$^{-1}$, $R_2(t_0)=2.35$ Gpc$^{-3}$ yr$^{-1}$ and $R_3(t_0)=3.29\times 10^{-2}$ Gpc$^{-3}$ yr$^{-1}$, as shown in Fig.~\ref{fig:mergertotal}.
It indicates that, in the monochromatic case, although the merger events of both $30~M_\odot-30~M_\odot$ PBH binaries and $60~M_\odot-30~M_\odot$ PBH binaries could occur at the same time, the major of merger events detected by LIGO-Vigo is the merger event of $30~M_\odot-30~M_\odot$ PBH binaries.
Therefore, in the monochromatic case, the effect of the merger history on the merger rate of PBH binaries is negligible.

\subsection{Power-law mass function}
In this subsection, we take the PBH mass function as a power-law form~\cite{Carr:1975qj}:
\m
\label{power}
 P(m)\approx {q-1\over M} \({m\over M}\)^{-q},
\n
with $500M\geq m\geq M$ and $q>1.5$. In the power-law case, we can rewrite~\eqref{mpbh}
and~\eqref{F(m)} as
\m
m_{\rm{pbh}}=M\frac{q}{q-1}\,,
\n
\m
F(m)={q \over m} \({m\over M}\)^{-q}\,
\n

\begin{figure}[htbp!]
\centering
\includegraphics[width = 0.48\textwidth]{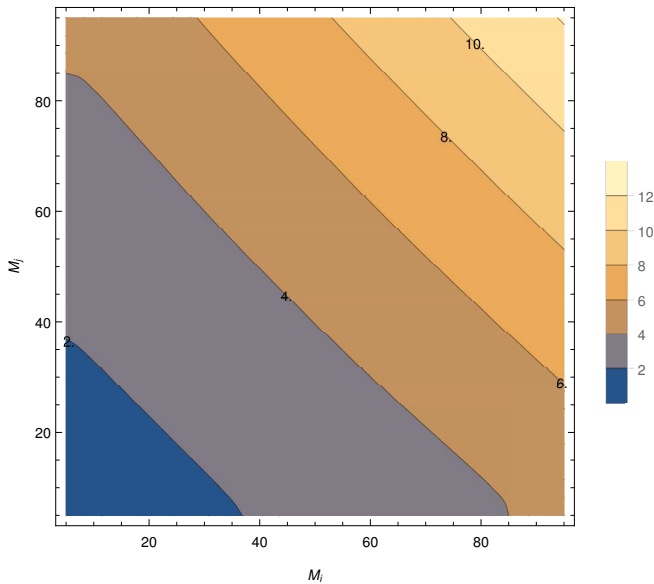}
\caption{\label{fig:Effects}
Contour of the ratio of the total merger rate density to the single-merger one in the PBH mass plane in the case of $f_{\rm pbh}=0.01$, $q=2.3$ and $M=0.2\Msun$.}
\end{figure}

Choosing $f_{\rm pbh}=0.01$, $q=2.3$, $M=0.2\Msun$, we can get $R_1(t_0)=9.66 \times 10^{3}$ Gpc$^{-3}$ yr$^{-1}$, $R_2(t_0)=1.15 \times 10^{2}$ Gpc$^{-3}$ yr$^{-1}$, $R_3(t_0)=5.00$  Gpc$^{-3}$ yr$^{-1}$. In power-law case,  the effect of the merger history on the merger rate of PBH binaries is small. However, the effect of the merger history on the merger rate density is significant in some region of the parameter space. For example, $\RR_1(t_0,30\Msun,30\Msun)=8.55 \times 10^{-7}$Gpc$^{-3}$ yr$^{-1}$$\Msun^{-2}$, $\RR_2(t_0,30\Msun,30\Msun)=8.90 \times 10^{-7}$Gpc$^{-3}$ yr$^{-1}$$\Msun^{-2}$, $\RR_3(t_0,30\Msun,30\Msun)=4.88 \times 10^{-8}$Gpc$^{-3}$ yr$^{-1}$$\Msun^{-2}$.
In Fig.~\ref{fig:Effects}, we show the ratio of the total merger rate density to the single-merger one in the PBH mass plane.
There are several gravitational wave events detected by LIGO-Virgo. Masses of black hole all are in  $(5 \Msun, 50 \Msun)$.  In such region, in the future, more and more coalescence events of black hole binaries will be detected by LIGO-Virgo~\cite{Wang:2019kaf,Chen:2019irf}. When we use the merger rate distribution to fit the mass function of PBH,  the effect of merger history on the merger rate density of PBH binaries can not be ignored.

\subsection{Log-normal mass function}
In this subsection, we take the PBH mass function as a log-normal form~\cite{Dolgov:1992pu,Green:2016xgy,Kuhnel:2017pwq}:
\m
\label{log}
 P(m) = \frac{1}{\sqrt{2 \pi} \sigma m} 
   \exp\(-\frac{\log^2(m/m_c)}{2 \sigma^2}\). 
\n
In the power-law case, we can rewrite~\eqref{mpbh}
and~\eqref{F(m)} as
\m
m_{\rm{pbh}}=m_c \exp(-\frac{\sigma^2}{2}).
\n
\m
F(m)=\frac{m_c}{\sqrt{2 \pi} \sigma m^2} 
   \exp\(-\frac{\sigma^2}{2}-\frac{\log^2(m/m_c)}{2 \sigma^2}\).
\n

\begin{figure}[htbp!]
\centering
\includegraphics[width = 0.48\textwidth]{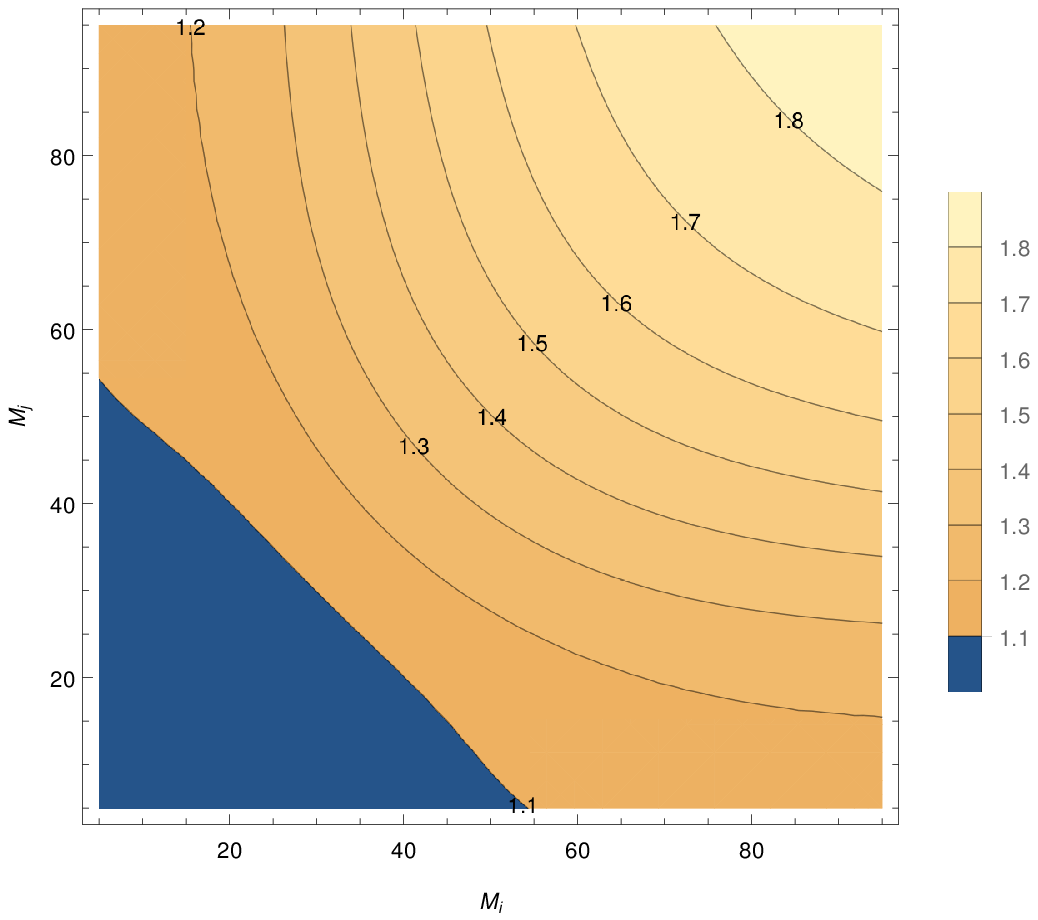}
\caption{\label{fig:Effects2}
Contour of the ratio of the total merger rate density to the single-merger one in the PBH mass plane in the case of $f_{\rm pbh}=0.01$, $m_c=15 \Msun$ and $\sigma=0.5$.}
\end{figure}

\begin{figure}[htbp!]
\centering
\includegraphics[width = 0.48\textwidth]{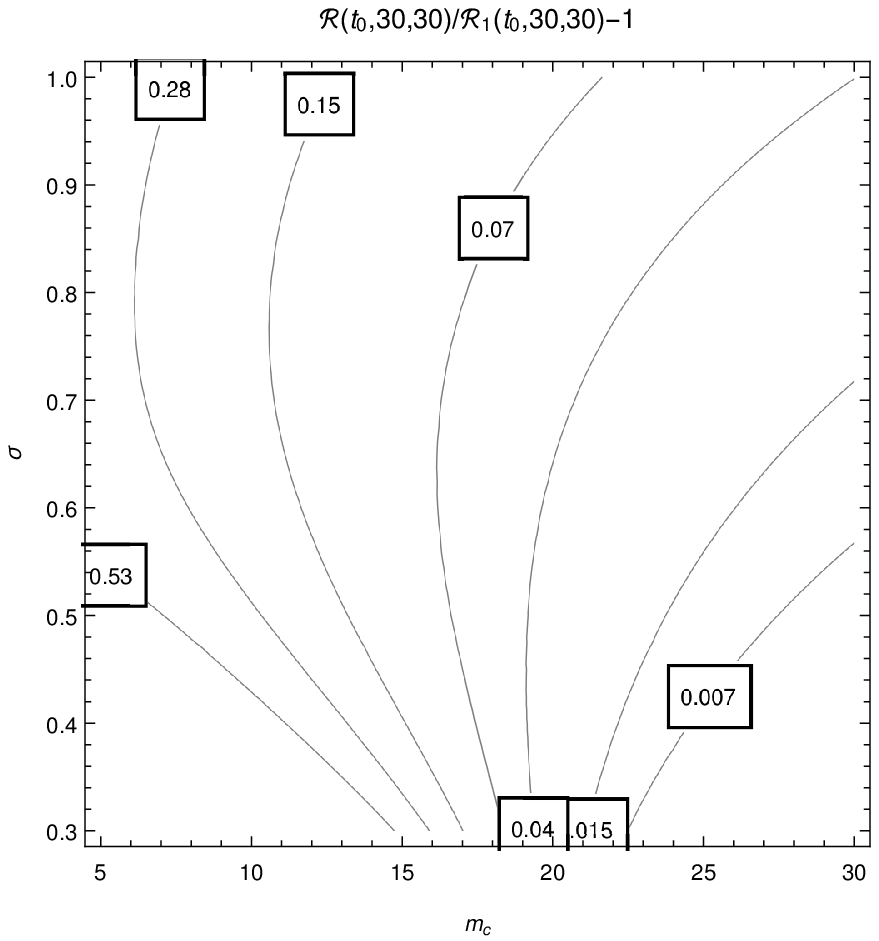}
\caption{\label{fig:Effects3}
Contour of $(\RR(t_0,30\Msun,30\Msun)/\RR_1(t_0,30\Msun,30\Msun)-1)$ in the parameter space of PBH mass function}
\end{figure}

Choosing $f_{\rm pbh}=0.01$, $m_c=15 \Msun$, $\sigma=0.5$, we can get $R_1(t_0)=423 $ Gpc$^{-3}$ yr$^{-1}$, $R_2(t_0)=6.5 $ Gpc$^{-3}$ yr$^{-1}$, $R_3(t_0)=0.1$  Gpc$^{-3}$ yr$^{-1}$. In log-normal case, the effect of the merger history on the merger rate of PBH binaries is also small. According to $\RR_1(t_0,30\Msun,30\Msun)=2.16 \times 10^{-2}$Gpc$^{-3}$ yr$^{-1}$$\Msun^{-2}$, $\RR_2(t_0,30\Msun,30\Msun)=2.14 \times 10^{-3}$Gpc$^{-3}$ yr$^{-1}$$\Msun^{-2}$, $\RR_3(t_0,30\Msun,30\Msun)=2.31 \times 10^{-5}$Gpc$^{-3}$ yr$^{-1}$$\Msun^{-2}$, the effect of the merger history on the merger rate density of PBH binaries could not be negligible in some region of the parameter space. In Fig.~\ref{fig:Effects2}, we show the ratio of the total merger rate density to the single-merger one in the PBH mass plane in the case of $f_{\rm pbh}=0.01$, $m_c=15 \Msun$ and $\sigma=0.5$. In Fig.~\ref{fig:Effects3}, we also plot the contour of $(\RR(t_0,30\Msun,30\Msun)/\RR_1(t_0,30\Msun,30\Msun)-1)$ in the parameter space of PBH mass function to show that the effect of the merger history on the merger rate density depend on the mass function.

\section{Conclusions}
\label{Conclusion}
We have developed the formalism to calculate the merger rate density of PBH binaries with a general mass function,
by taking into account the merger history of PBHs.
In the monochromatic case, we find that $R_1 \gg R_2  \gg R_3$, which is independent on $f_{\rm pbh}$.
Therefore, the effect of the merger history on the merger rate of PBH binaries is negligible.
However, the multiple-merger events may play an important role in the merger rate density of PBH binaries in the non-monochromatic case.
For example, for the power-law and log-normal mass function, the effect of the merger history on the merger rate density of PBH binaries could not be negligible.
In the future, more and more coalescence events of black hole binaries will be detected by LIGO-Virgo.
This will provide more rich information on the merger rate distribution of black hole binaries to test the PBH scenario.

We calculate the merger rate density of PBH binaries up to three mergers.
In principle, one can directly calculate it at more than three mergers by using the formalism developed in the present paper.
Since the contribution of the merger history on the merger rate density of PBH binaries
depends on the mass function and the mass region,
it is hard to judge whether mergers of higher order should be computed for a generic mass function.

The effects of the tidal field from the smooth halo, the encountering with other PBHs, the baryon accretion and  present-day halos, are carefully investigated in~\cite{Ali-Haimoud:2017rtz}.
It is found in~\cite{Ali-Haimoud:2017rtz} that these effects make no significant contributions to the overall merger rate. We therefore neglected these subdominant effects throughout our computation.

\begin{acknowledgments}
This work is supported in part by the National Natural Science Foundation of China Grants
No.11575272, No.11435006, No.11690021, No.11690022, No.11851302 and No.11821505,
in part by the Strategic Priority Research Program of the Chinese Academy of Sciences Grant No. XDB23030100,
No. XDA15020701 and by Key Research Program of Frontier Sciences, CAS.
\end{acknowledgments}


\bibliography{merger_v11}

\end{document}